\documentclass[sigconf]{acmart} 

\AtBeginDocument{%
  }

\setcopyright{acmlicensed}
\copyrightyear{2026}
\acmYear{2026}

\acmConference[CHI '26]
{Cross-Sensory Futures: Rewiring Perception in HCI, Workshop at the 2026 CHI Conference on Human Factors in Computing Systems}
{April 13--17, 2026}
{Barcelona, Spain}

\begin{document}

\title[Rewiring Perceived Doability in VR]{Rewiring Perceived Doability in VR: Hand Redirection as a Subtle Cross-Sensory Support for Sustained Practice}

\author{Isidro {Butaslac}~III\hspace{0.1em}}
\authornote{Corresponding author.}
\authornote{Nara Institute of Science and Technology (NAIST).}
\orcid{0000-0003-1172-7611}
 \affiliation{%
  \institution{NAIST}
  \city{Ikoma}
  \country{Japan}
  }
\email{isidro.b@naist.ac.jp}

\author{Yota Nagaya}
 \affiliation{%
  \institution{NAIST}
  \city{Ikoma}
  \country{Japan}
  }
\email{nagaya.yota.oz5@naist.ac.jp}

\author{Almira Princess Redoble}
 \affiliation{%
  \institution{Ateneo de Manila University}
  \city{Quezon City}
  \country{Philippines}
  }
\email{almira.redoble@student.ateneo.edu}

\author{Jordan Aiko {Deja}} 
\orcid{0000-0001-9341-6088}
 \affiliation{%
  \institution{De La Salle University}
  \city{Manila}
  \country{Philippines}
  }
\email{jordan.deja@dlsu.edu.ph}

\author{Nicko R. {Caluya}} 
\orcid{0000-0003-4924-958X}
 \affiliation{%
  \institution{Ritsumeikan University}
  \city{Osaka}
  \country{Japan}
  }
\email{nicko@fc.ritsumei.ac.jp}

\author{Maheshya Weerasinghe}
\orcid{0000-0003-2691-601X}
\affiliation{%
  \institution{University of Primorska}
  \city{Koper}
  \country{Slovenia}
 }
\email{maheshya.weerasinghe@famnit.upr.si}

\author{Taishi Sawabe}
\orcid{https://orcid.org/0000-0001-9244-479X}
 \affiliation{%
  \institution{NAIST}
  \city{Ikoma}
  \country{Japan}
  }
\email{t.sawabe@is.naist.jp}

\author{Hirokazu Kato}
\orcid{https://orcid.org/0000-0003-3921-2871}
 \affiliation{%
  \institution{NAIST}
  \city{Ikoma}
  \country{Japan}
  }
\email{kato@is.naist.jp}

\author{Eric Cesar {Vidal} Jr.}
 \affiliation{%
  \institution{Ateneo de Manila University}
  \city{Quezon City}
  \country{Philippines}
  }
\email{evidal@ateneo.edu}

\renewcommand{\shortauthors}{Butaslac et al.}

\begin{abstract}
In everyday life, physical effort is often minimized and convenience is prioritized, making it difficult for many people to sustain light exercise and stretching despite well-known long-term benefits. This challenge often arises not from objective movement limitations, but from whether an action feels \emph{doable} in the moment and, therefore worth continuing. This position paper argues that subtle VR hand redirection (HR) can be reframed as a form of cross-sensory support for sustained practice by targeting \emph{perceived doability}: a moment-to-moment cognitive appraisal that an action is within one’s capability while requiring manageable effort. We propose that conservative HR, applied within known perceptual limits, can create repeated micro-success experiences (e.g., reaching a virtual goal earlier with similar physical movement). These micro-successes may increase continuation intention and early re-engagement without relying on overt pressure or intensive coaching. At the same time, such support raises questions about autonomy and authenticity. We therefore articulate two research questions: (RQ1) how HR shifts perceived doability to support sustained practice and positive behavior change; and (RQ2) when HR functions as acceptable support versus becoming counterproductive by undermining authenticity, agency, trust, or fostering dependence. We present an initial sit-and-reach VR prototype, outline a research plan, and identify key design tensions to spark community discussions on autonomy-preserving cross-sensory futures in HCI.
\end{abstract}



\maketitle

\section{Introduction}

\begin{figure*}[ht]
  \centering
  \includegraphics[width=\linewidth]{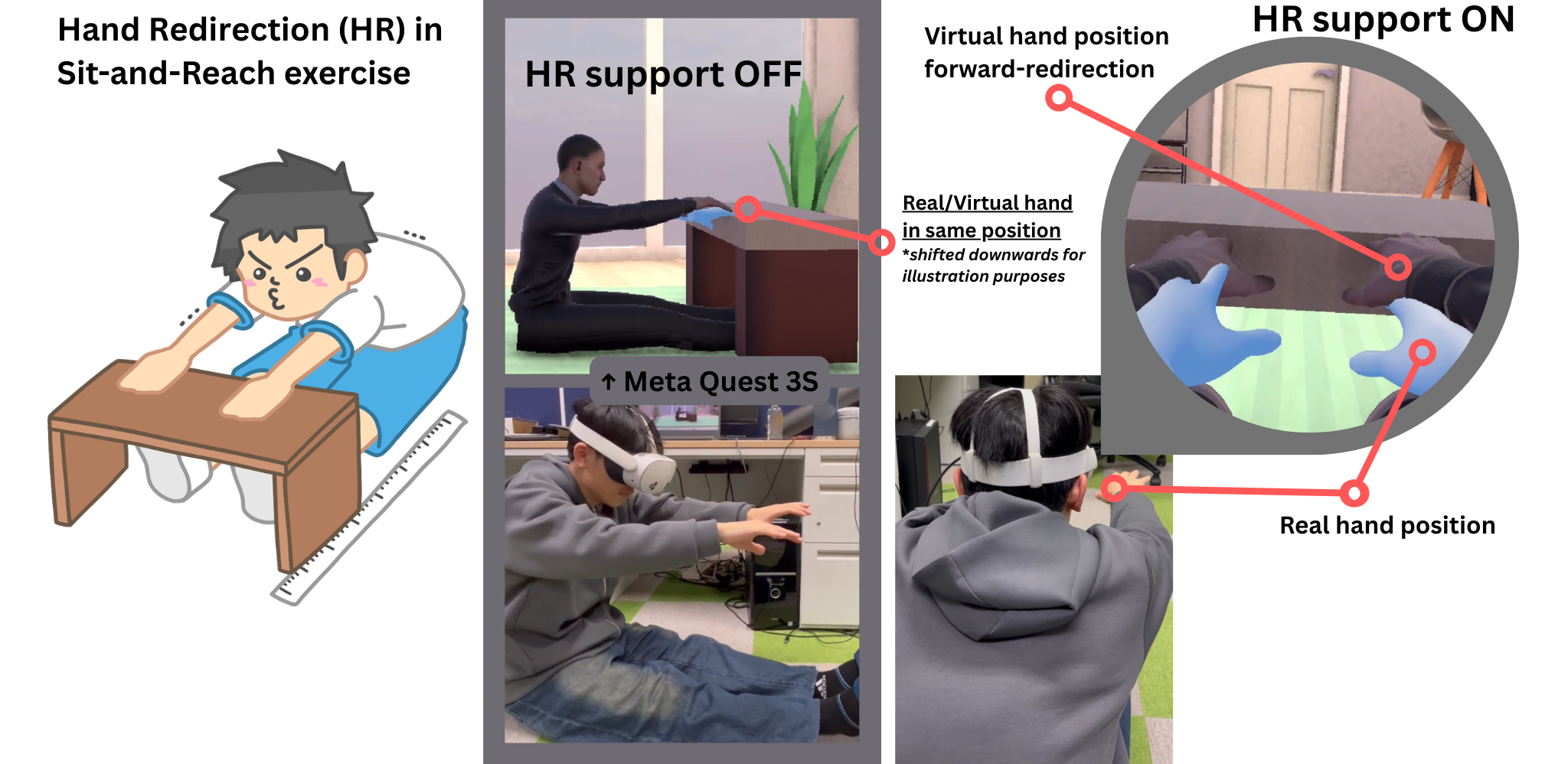}
  \caption{Our initial Meta Quest 3S prototype of HR as subtle cross-sensory support in a sit-and-reach stretching task. This illustrates the intended visuo--proprioceptive remapping between real and virtual hand positions.}
  \Description{VR Hand Redirection prototype and teaser image of big picture idea of this research.}
  \label{HR-teaser}
\end{figure*}

Repetitive light exercise and stretching are widely recommended for maintaining mobility, reducing pain, and supporting independent living. However, many people still struggle to sustain them \cite{Clawson2015NoLongerWearing,Epstein2016BeyondAbandonment}. The familiar pattern is: know it is good but cannot keep doing it. This intention--continuation gap is especially visible in older adults, for whom even simple flexibility routines can feel effortful, discouraging, or easy to postpone despite a clear long-term value \cite{McAuley1993LongTermMaintenanceExercise,McAuley2011SelfEfficacyImplications}. Similar dynamics apply to people starting a new routine or returning after a long break, when early sessions can feel disproportionately effortful and progress is slow or ambiguous. A central obstacle is not necessarily the objective difficulty of the movement, but the moment-to-moment experience of whether it feels doable and worth persisting in \cite{McAuley2011SelfEfficacyImplications,SilvaSmith2024PhysicalActivityInterventions}. In this research work, we argue that augmented or virtual reality (AR/VR) can support sustained practice by targeting this subjective pivot point: \emph{perceived doability}, defined here as the self's cognitive appraisal of capability. It is the ongoing reinterpretation that an action is within one’s capability, demands manageable effort, and is therefore worth continuing. Within the scope of this research, we treat perceived doability as a momentary, task-specific appraisal that integrates expected effort, perceived progress, and confidence to continue. Importantly, in our title we use ``rewiring'' as a metaphor for cognitive reframing of this appraisal, not as a claim about long-term neural change or effects.

We situate this approach within the workshop’s cross-sensory framing by treating \textbf{hand redirection (HR)} as an instance of cross-sensory interaction, namely a controlled visuo-proprioceptive remapping in which visual evidence of hand position is subtly altered relative to proprioception \cite{Zenner2019EstimatingDetectionThresholds,Hartfill2021AnalysisDetectionThresholdsHandRedirection}. Rather than using remapping primarily to extend reach or solve geometric constraints, we explore it as subtle cross-sensory support that can reshape what users feel their body can achieve in the moment. In cross-sensory terms, we focus on how small changes in visual evidence can recalibrate proprioceptive inference and, importantly, the cognitive appraisal that governs persistence. This may shift perceived doability and by extension, continuation in repetitive practice. Fig. \ref{HR-teaser} illustrates our prototype concept in a sit-and-reach style stretching task using a head-mounted display (Meta Quest 3S). In an HR support OFF condition, the virtual and real hands remain aligned. In an HR support ON condition, the virtual hand is forward-redirected relative to the real hand, enabling the user to reach a target earlier in VR while performing a similar physical movement. This creates a controlled discrepancy that can potentially increase perceived doability by amplifying perceived progress and reducing perceived effort. But at the same time, it raises concerns about agency, trust, and dependence when assistance becomes perceptible or attribution becomes ambiguous \cite{Liao2025FocusDrivenAugmentedFeedback}.

Within this background, we ask two research questions that connect cross-sensory remapping to sustained practice outcomes while foregrounding acceptability boundaries.

\textbf{RQ1:} How can subtle visuo–proprioceptive remapping via HR in VR shift users’ cognitive appraisal of capability (perceived doability) to support sustained practice and positive behavior change?

\textbf{RQ2:} Under what redirection parameters and design strategies does HR function as acceptable cross-sensory support (improving self-efficacy and continuation) versus becoming counterproductive (reducing agency/trust, creating dependence)?

\section{Related Work}

\subsection{Self-efficacy, adherence, and habit formation}

Sustained practice is often constrained less by whether an intervention is ``effective'' in principle than by whether people continue to use it in everyday life. Previous HCI work in personal informatics shows that dropout is common and is often driven by factors that extend beyond immediate results: expectation mismatch, declining motivation, and life changes can make tools stop fitting user-evolving routines and priorities \cite{Clawson2015NoLongerWearing}. Moreover, discontinuation is not always a straightforward failure case. Epstein et al. argue that ``abandonment'' can reflect burden/cost, discomfort, privacy concerns, shifting life circumstances, or even a sense of having learned enough, and they describe how people interpret life afterwards (e.g., guilt or frustration, but also relief and continued use of learned skills) \cite{Epstein2016BeyondAbandonment}. Together, these perspectives motivate treating adherence and tool drop-off as a design problem: systems must remain compatible with messy, changing practices rather than assuming continuous, uniform engagement \cite{Clawson2015NoLongerWearing,Epstein2016BeyondAbandonment}.

Within health behavior and physical activity research, social-cognitive accounts highlight self-efficacy as a key determinant of whether people initiate and maintain behavior over time. McAuley et al. identify exercise self-efficacy as a strong predictor of adherence and maintenance at follow-up after structured exercise programming \cite{McAuley1993LongTermMaintenanceExercise}. More broadly, they argue that self-efficacy is a central, modifiable determinant shaping effort, persistence, and long-term maintenance, with mastery experiences emphasized as the most potent source of efficacy beliefs \cite{McAuley2011SelfEfficacyImplications}. Extending this to intervention design for long-term physical activity, Silva-Smith et al.’s review of the Health Action Process Approach (HAPA) synthesizes how self-efficacy interacts with planning, coping, and recovery processes to bridge the intention--behavior gap \cite{SilvaSmith2024PhysicalActivityInterventions}. This body of work supports a core implication for adherence-focused systems: sustaining practice depends on users’ appraisals of capability and manageability, not solely on objective performance or physiological change \cite{McAuley1993LongTermMaintenanceExercise,McAuley2011SelfEfficacyImplications,SilvaSmith2024PhysicalActivityInterventions}.

Recent AR/VR and interactive-system studies further suggest that efficacy beliefs can be shaped by how ``success'' is experienced and remembered, even when raw performance is held constant. Inamura et al. manipulates virtual success experiences in a VR motor-skill task (e.g., vicarious experience ordering, rewards, and difficulty adjustment) and show that self-efficacy is closely tied to remembered and anticipated success rather than objective task outcomes \cite{Inamura2025EnhancingSelfEfficacyVRKendama}. From an HCI design perspective, systems have also framed behavior modification as something that can be scaffolded when self-management is difficult; for example, TSUNDERE Interaction uses adaptive feedback shaped by user state, drawing on operant-conditioning-style reinforcement to sustain motivation \cite{Tainaka2021TSUNDEREInteraction}. Collectively, these works motivate interventions that do not merely deliver instructions or feedback, but actively support continuation by shaping the perceived attainability of action through experience design \cite{Inamura2025EnhancingSelfEfficacyVRKendama,Tainaka2021TSUNDEREInteraction}.

\subsection{AR/VR for Motivation and Engagement}

Building on the evidence that continuation hinges on users’ appraisals of attainability and effort (rather than objective outcomes alone), AR/VR systems have been widely explored as tools for sustaining motivation and engagement during repetitive practice \cite{Butaslac2023SystematicReviewARTraining}. In rehabilitation and exercise contexts, this work often leverages immersive feedback, game mechanics, and structured programs to make training feel more enjoyable, meaningful, and worth returning to across sessions. For example, VR serious games for hand and finger rehabilitation have been iteratively designed to encourage effort and motivation, with multi-session deployments reporting increases in training activity and engagement beyond one-off novelty effects \cite{Bressler2024VRSeriousGameHandRehab}. Similarly, comparative studies of enriched versus non-enriched VR rehabilitation environments show that design choices affecting context and feedback can increase intrinsic motivation (e.g., interest/enjoyment) and engagement indicators such as time spent and in-task performance \cite{Zuki2024AssessingImpactEnrichedVR}. Most AR/VR adherence designs primarily operate through explicit motivational scaffolds (game mechanics, feedback, rewards) or structured programs. Less explored is whether changing the sensorimotor evidence of progress can shift the appraisal that determines whether users persist.

A parallel line of work in VR exergames and fitness systems further demonstrates how sustained practice can be scaffolded through design for engagement and adaptation. Cmentowski et al. provide a detailed account of how VR training systems can be designed and evaluated around motivation/engagement while also accounting for safety and early signals relevant to longer-term participation \cite{Cmentowski2023NeverSkipLegDay}. Longitudinal designs, such as LightSword explicitly targets longer-term training via a multi-session program with follow-up months later, reporting sustained benefits alongside motivation and compliance factors \cite{Du2024LightSword}. Related longitudinal evidence suggests that specific VR design interventions (e.g., self-competition mechanics) can shift adherence-relevant outcomes such as intrinsic motivation, flow/absorption, self-efficacy, and intention for future use \cite{Michael2020RaceYourselves}. Beyond exercise, VR experiences that integrate physiological and haptic feedback can also modulate engagement-adjacent affective states, as shown by Hype Live, where biometric-linked sensory feedback improved participants’ sense of unity \cite{Abe2022HypeLiveBiometricFeedback}. Taken together, these studies suggest that immersive experiences can influence motivational and affective dynamics that matter for continuation, not only task performance \cite{Abe2022HypeLiveBiometricFeedback,Bressler2024VRSeriousGameHandRehab,Cmentowski2023NeverSkipLegDay,Du2024LightSword,Michael2020RaceYourselves,Zuki2024AssessingImpactEnrichedVR}.

However, AR/VR designs that aim to sustain engagement must balance the trade-offs between enjoyment and effort and between assistance and autonomy. The choices for interaction techniques can shift perceived workload and acceptance; for example, controller-free hand input in AR can increase enjoyment and acceptance but also increase workload/fatigue and introduce reliability issues relative to controller-based interaction \cite{Haneling2025BeyondControllers}. Similarly, adaptive assistance in VR rehabilitation can help maintain focus and engagement, but only when carefully calibrated, as over-assistance risks reducing the sense of agency and engagement \cite{Liao2025FocusDrivenAugmentedFeedback}. Adaptive VR fitness systems likewise highlight how engagement outcomes (e.g., flow, enjoyment, motivation) co-evolve with perceived exertion and feedback over repeated sessions \cite{MartinNiedecken2019ExerCubeVsPersonalTrainer}. These findings foreground a recurring design tension for sustained practice, that is, interventions must make action feel more manageable while preserving users’ sense that the activity remains genuinely theirs \cite{Haneling2025BeyondControllers,Liao2025FocusDrivenAugmentedFeedback,MartinNiedecken2019ExerCubeVsPersonalTrainer}.

This tension motivates interest in mechanisms that can support practice by shaping the ``experienced evidence'' users draw on when appraising capability and effort, potentially without relying solely on explicit rewards, instructions, or overt feedback. In rehabilitation contexts, approaches that subtly alter the sensorimotor experience have been suggested to be promising for increasing motivation and willingness to continue. For example, Xiong et al. report that a visuo--proprioceptive mismatch technique can increase motivation and effort while remaining largely unnoticed and acceptable to participants in the context of longer-term rehabilitation programs \cite{Xiong2024ToReachTheUnreachable}. This opens a pathway to consider HR, not merely as an interaction technique, but as a form of cross-sensory support that may shift perceived doability while raising critical questions about acceptability, agency, and trust.

\subsection{HR as a Cross-Sensory Support Tool}

Following the motivation to shape the experiential evidence users rely on when appraising capability and effort, HR offers a concrete cross-sensory mechanism: it introduces a controlled visuo--proprioceptive remapping by subtly altering how the virtual hand is rendered relative to the physical hand. A substantial body of work has characterized when such remapping remains unnoticed, feels embodied, and stays safe for interaction, providing essential foundations for treating HR as a ``support tool'' rather than a disruptive illusion. Conservative psychophysical measurements have established detection thresholds for desktop-scale HR across multiple warping styles, yielding practical bounds for ``subtle'' remapping that can be leveraged without immediate awareness \cite{Zenner2019EstimatingDetectionThresholds}. Subsequent analyses further quantify detection thresholds for gain-based HR across movement axes and directions, reinforcing that perceptual limits are structured (i.e., not uniform across motions) and therefore must be respected when targeting a ``natural-feeling'' interaction \cite{Hartfill2021AnalysisDetectionThresholdsHandRedirection}.

Beyond detectability, embodiment-related factors shape the way redirection is experienced. Feick et al. examine detectability under different avatar completeness levels and report that more complete avatars can increase embodiment while not producing practically relevant changes in conservative detection thresholds \cite{Feick2024ImpactAvatarCompleteness}. This distinction matters for cross-sensory support. Even if the ``unnoticed'' range remains similar, the felt credibility of the virtual body can change how users interpret success and effort within that range \cite{Feick2024ImpactAvatarCompleteness}. Taken together, these findings suggest that HR can be engineered to remain subtle while still operating within an embodied interaction context, making it a plausible candidate for shifting users’ cognitive appraisal of capability (\textbf{RQ1}) without overt instructional persuasion. Mechanistically, sub-threshold HR can amplify perceived progress (reaching the target sooner) and reduce perceived effort-to-success ratio, creating ``micro mastery experiences'' that may update capability appraisal even when objective movement is similar.

Crucially, the acceptability of HR is not solely a property of a fixed offset or gain, it is shaped by feedback context and attentional conditions. Tanaka et al. show that detection thresholds differ between real-time discrepancies and replay-based discrepancies, with replay context shifting psychophysical markers, such as just-noticeable differences (JND), when redirection is applied during movement \cite{Tanaka2025DetectionThresholdsForReplay}. Complementing this, Li et al. model noticeability using gaze behavior and demonstrate that environmental visual stimuli systematically modulate detectability, motivating context-aware control rather than assuming a universally ``safe'' setting \cite{Li2025ModelingImpactVisualStimuli}. These results directly support the \textbf{RQ2} framing, that is, whether HR functions as acceptable cross-sensory support depends on both \emph{how} remapping is applied and \emph{what} perceptual context users are in, with implications for trust and perceived transparency and attribution ambiguity if users notice inconsistencies \cite{Li2025ModelingImpactVisualStimuli,Tanaka2025DetectionThresholdsForReplay}.

Finally, safety and contact synchrony considerations highlight that design choices influence both subjective experience and interaction integrity. Zhou and Popescu compare redirection mechanisms aimed at safe interaction and contact alignment, showing that algorithmic strategies (e.g., dynamic versus static/no redirection) can affect synchronization outcomes and artifact profiles that resemble noticeability-related breakdowns \cite{Zhou2025DynamicRedirectionForSafeInteraction}. For our purposes, this reinforces that ``acceptable support'' is not merely about staying below a detection threshold. More importantly, it also requires maintaining interaction coherence and safeguarding embodied expectations. In this light, HR can be positioned as a form of cross-sensory support whose promise lies in subtly reshaping perceived doability (\textbf{RQ1}), while whose risks center on boundary conditions where subtlety fails and agency/trust are undermined (\textbf{RQ2}). This implies that beyond tuning offsets/gains, HCI researchers must consider strategies for calibration, user control, and adaptive/fade-out policies that preserve agency while delivering support.

\section{HR for Rewiring Perceived Doability}

\subsection{Position of this Research}
The stance of this research is that HR can function as subtle cross-sensory support for sustained practice by targeting a specific psychological bottleneck: perceived doability. As mentioned earlier, we have defined perceived doability as a person's cognitive appraisal of capability, namely the moment-to-moment reinterpretation that an action is within one’s capability while requiring manageable effort, and therefore it is worth persisting in. In light exercise and stretching, this appraisal is often the difference between continuing and dropping out, especially when routines are repetitive and progress is slow or ambiguous \cite{McAuley1993LongTermMaintenanceExercise,McAuley2011SelfEfficacyImplications,SilvaSmith2024PhysicalActivityInterventions}.

To address \textbf{RQ1}, we hypothesize that small visuo--proprioceptive remapping via HR can reliably shift perceived doability through a success-experience pathway. By slightly increasing the apparent reach or progress in VR while the user performs a similar physical movement, HR can create repeated micro-confirmations of ``I could do it'' without requiring large changes in task structure or explicit motivational messaging. This mechanism is conceptually aligned with prior evidence that moderate, non-realistic movement mapping can increase perceived competence and preserve naturalness. For example, Granqvist et al.\ show that exaggerating avatar flexibility in VR within a moderate range can improve perceived competence and naturalness, suggesting that carefully tuned remapping can reshape how capable an action feels without necessarily breaking immersion \cite{Granqvist2018ExaggerationAvatarFlexibility}. Translating this into stretching practice, like the example in Fig.~\ref{HR-teaser}, we treat HR as a cross-sensory cue that adjusts the experiential evidence users draw on when evaluating capability and effort. If successful, this shift in perceived doability should increase continuation intention and early re-engagement, providing a plausible pathway toward sustained practice in the long-term.

To address \textbf{RQ2}, we adopt the working hypothesis that agency, trust, and autonomy can be preserved by deliberately managing user awareness of HR rather than attempting to eliminate it entirely. Specifically, while HR operates as subtle cross-sensory support during action to enhance perceived doability in the moment, we posit that complementary mechanisms outside the immediate HR experience (e.g., session-level progress visualization, explicit feedback about actual performance trends, or gradual fading of support) can make users aware of their ``true capability'' trajectory over time. This layered approach is intended to mitigate overdependence by ensuring that perceived success remains attributable to the self and oriented toward improvement, rather than masking limitations or encouraging stagnation.

Together, these working hypotheses directly address our research questions. \textbf{RQ1} asks whether HR-driven visuo--proprioceptive remapping can shift cognitive appraisal of capability (perceived doability) to support sustained practice, while \textbf{RQ2} asks when such support remains acceptable versus becoming counterproductive by undermining agency, trust, or autonomy. The core investigation is therefore not simply ``better performance in VR,'' but whether HR can durably increase willingness to return to practice while preserving a credible sense of self-authored action and achievement.

\subsection{Research Plan}

We outline our research plan as follows. The target application is forward-bend seated stretching in a sit-and-reach style task using our initial Meta Quest 3S prototype (Fig.~\ref{HR-teaser}). The population of interest includes individuals who do not engage in regular exercise or stretching and who therefore often experience low perceived doability when initiating light physical routines. Older adults are treated as an exemplar subgroup within this broader population, as light stretching is commonly recommended for them yet adherence to regular practice is frequently difficult \cite{McAuley1993LongTermMaintenanceExercise}. The core experimental contrast is HR support ON versus HR support OFF, with short repeated exposure designed to investigate immediate appraisal changes and near-term continuation intentions. This research plan was approved by the Ethics Committee of Nara Institute of Science and Technology (2025-I-51).

\textbf{Design logic.} Participants complete repeated sessions in VR under HR OFF (veridical hand alignment) and HR ON (forward redirection), counterbalanced when appropriate. We focus on brief sessions that are long enough to observe changes in appraisal and engagement but short enough to manage fatigue and discomfort. To connect to sustained practice rather than one-session novelty, we plan to include a subjective follow-up questionnaire that checks whether participants would continue the routine and whether they actually re-engage when given the opportunity.

\textbf{Measures and outcomes.}
To test the proposed mechanism (\textbf{RQ1}), we will assess perceived doability directly through self-report items that target capability appraisal and manageable effort, along with exercise self-efficacy, perceived effort or strain, and continuation intention. To ensure applicability of the measures and capture a key constraint in sustained practice, we will also measure VR-induced fatigue and discomfort. In this context, a validated multidimensional fatigue instrument for VR provides a principled way to quantify fatigue across general, visual, motivational, and emotional dimensions, and to account for factors such as session duration and experience \cite{CintoraSanz2026VirtualMixedRealityFatigueQuestionnaire}. To evaluate engagement in a rehabilitation-relevant manner, we will incorporate a validated engagement rating approach where feasible. The Hopkins Rehabilitation Engagement Rating Scale - Reablement Version (HRERS-RV) offers a concrete engagement metric that has been shown to predict functional outcomes and can serve as a pragmatic indicator of whether HR support translates into meaningful participation \cite{Mayhew2019HopkinsRehabilitationEngagementRatingScale}.

To address boundary conditions (\textbf{RQ2}), we will collect acceptability and relevant responses focused on perceived agency, trust, and dependence risks (i.e., the reduced willingness/confidence to perform the task without HR after exposure and/or a strong preference to practice only when support is enabled). This includes perceived authenticity of success (did I do it or did the system do it for me), perceived transparency and attribution ambiguity (did the system ``steer'' my success?), and willingness to use in real exercise contexts. We treat these commentaries as an important part in the evaluation because HR support is only valuable if it remains acceptable and does not undermine intrinsic motivation.

\textbf{What we expect to learn.}
For \textbf{RQ1}, we will test whether HR support ON produces a measurable shift in perceived doability and whether that shift is associated with continuation intention and early re-engagement. A key analytic question is whether perceived doability mediates the relationship between HR condition and continuation-related outcomes, consistent with the proposed mechanism. For \textbf{RQ2}, we will map where support remains acceptable versus where it becomes counterproductive, using patterns in agency, trust, fatigue, and acceptability responses to identify practical boundaries. The overall outcome is not a finalized parameter prescription, but a principled empirical basis for designing HR as a cross-sensory support that improves self-efficacy and continuation while avoiding agency and trust failures.

\section{Key Tensions and Open Questions}
In the following, we outline three tensions intended for the workshop discussion and sharpen design/research agendas.

\subsection{Assistance vs. Authenticity}
HR-based support can make progress feel more attainable by enabling users to reach a virtual goal earlier while performing a similar physical movement. This directly motivates an authenticity question: \emph{did I succeed or did the system author my success?} The concern is not merely philosophical; authenticity can shape whether the experience updates capability appraisal in a stable way or is discounted as ``the system helping.'' As Inamura et al. showed, system-shaped \emph{virtual success experiences} can increase self-efficacy even when objective performance gains are not proportional, highlighting how remembered/anticipated success can be influenced by experience design \cite{Inamura2025EnhancingSelfEfficacyVRKendama}. HR may operate similarly, as it can create repeated ``micro mastery'' signals, but the psychological effect may hinge on attribution (self-caused vs.\ system-caused).

\textbf{Discussion Points}
\begin{itemize}
  \item \textbf{Attribution design:} What cues make HR-mediated success feel self-authored rather than externally granted (e.g., emphasizing effort consistency, showing gradual progress rather than abrupt ``wins'')?
  \item \textbf{Employing authenticity:} What should we measure beyond self-report (e.g., persistence when HR is removed; willingness to practice without VR; confidence in real-world transfer)?
  \item \textbf{Ethical stance:} Is it acceptable to intentionally increase perceived success when the mapping is non-veridical (correspondence to reality is not 1:1), if the goal is sustained practice and positive habits? Under what disclosure or consent models is it justifiable?
\end{itemize}

\subsection{Agency/Trust vs. Subtlety}
A core advantage of HR is that it can remain \emph{subtle} (i.e., below conservative detection thresholds) and thereby avoid breaking immersion or imposing overt motivational messaging. Yet subtlety creates a second tension: \emph{if users do not notice, what protects agency and trust?} A system can be ``unnoticed'' and still be consequential. Feick et al.\ report that prolonged exposure to unnoticeable HR can degrade proprioceptive accuracy with limited immediate recovery, motivating concerns that cross-sensory support may introduce latent sensorimotor costs even when the experience feels natural \cite{Feick2025DelusionizedPotentialHarms}. This tension is especially relevant in sustained practice contexts; the intended benefit is repeated use, but repeated exposure is exactly what may accumulate risk (perceptual drift, safety concerns, or post-session uncertainty about one’s own capability).

\textbf{Discussion Points}
\begin{itemize}
  \item \textbf{What counts as acceptable manipulation?} Should ``subtle'' imply ``safe,'' or do we need separate criteria (e.g., bounded aftereffects, reversible recalibration, no increase in real-world risk)?
  \item \textbf{Transparency strategies:} When does disclosure preserve trust without collapsing the effect? Are there middle-ground designs (e.g., ``support mode'' framing, user-adjustable intensity, contextual consent)?
  \item \textbf{Designing for boundaries:} How should systems respond when subtlety fails (user notices mismatch, discomfort rises, trust drops)? Should HR be adaptive to attention/gaze or fatigue signals, and what is the ethical cost of such adaptation?
\end{itemize}

\subsection{Dependence vs. Empowerment}
Even if HR increases perceived doability in-session, a third tension asks whether such support \emph{empowers} users to practice independently or instead encourages reliance on the system. This maps to the guidance hypothesis in motor learning. Strong guidance can improve performance during practice but harm retention/transfer when learners come to depend on it, although reducing or fading guidance can mitigate these effects \cite{Winstein1994EffectsPhysicalGuidanceMotorLearning}. In AR training more broadly, we have distinguished \emph{support} from \emph{training} systems: support-oriented systems may help users accomplish tasks now but risk reducing independent capability unless they are deliberately designed toward autonomy \cite{Butaslac2023SystematicReviewARTraining}. Translating this to HR, a doability-boost remap could be beneficial in the early stages, yet counterproductive if users become unwilling to practice without it or if confidence drops when support is removed.

\textbf{Discussion Points}
\begin{itemize}
  \item \textbf{Fading and ``exit ramps'':} Should HR be designed to fade over sessions (or within-session) to encourage autonomy? What fading schedules preserve motivation without triggering frustration or dropout?
  \item \textbf{Measuring dependence:} What behavioral markers indicate harmful reliance (e.g., refusal to practice in HR OFF; reduced self-efficacy in non-assisted conditions; narrowed willingness to exercise outside VR)?
  \item \textbf{Cross-sensory empowerment:} Can cross-sensory support be designed to \emph{teach} users a more confident appraisal (e.g., recalibrating perceived effort/progress) that persists beyond VR, rather than merely providing a crutch?
\end{itemize}

\noindent
Across these tensions, the main point for discussion is to treat HR as a concrete case for broader cross-sensory futures: how do we design interactions that intentionally reshape perception and appraisal while preserving authenticity, agency, and long-term autonomy? These questions refine our \textbf{RQ1--RQ2} agenda from ``does it work?'' to ``when is it desirable, acceptable, and empowering?''.

\section{Conclusion}

This position paper argues that HR can be reframed from an interaction technique into a form of \emph{subtle cross-sensory support} for sustained practice in repetitive, low-intensity exercise (e.g., sit-and-reach stretching). We introduce \emph{perceived doability} as a psychological bottleneck underlying the intention--continuation gap: a moment-to-moment appraisal that an action is within one’s capability, demands manageable effort, and is therefore worth persisting in. By applying conservative, safe visuo--proprioceptive remapping in VR, HR may create repeated micro-success experiences that amplify perceived progress, potentially increasing continuation intention and early re-engagement, but may also raise critical questions about authenticity, agency, trust, and dependence.

Guided by the workshop’s theme of \emph{cross-sensory interaction} (where stimulation in one modality reshapes perception, cognition, or experience in another) we pose two research questions: \textbf{RQ1} asks how subtle visuo--proprioceptive remapping via HR can ``rewire'' perceived doability to support sustained practice and positive behavior change. \textbf{RQ2} asks under what redirection parameters and design strategies HR functions as an acceptable cross-sensory support versus becoming counterproductive by reducing agency/trust, creating dependence, or undermining authenticity. Together, these questions position HR as a concrete case for workshop discussion about when perceptual remapping should be treated as supportive experience design and when it crosses ethical or practical boundaries.

Our near-term future work is to empirically test HR support ON versus OFF in a sit-and-reach VR prototype, measuring perceived doability, self-efficacy, effort, fatigue, engagement, and continuation-related outcomes, alongside acceptability indicators such as attribution, transparency, and trust. Beyond this initial study, our research vision is to develop design principles for \emph{autonomy-preserving cross-sensory support}. That is, the adaptive and context-aware remapping that stays within conservative perceptual bounds, including ``exit ramps'' (i.e., fading support policies) to mitigate dependence, and clarifying accountability so users retain a credible sense of self-authored progress. More broadly, we aim to contribute a cross-sensory perspective on behavior change technologies, not only asking whether assistance increases performance in-the-moment, but whether it reshapes appraisals in ways that sustain practice while remaining acceptable, safe, and empowering.

\section*{Declaration on Generative AI}
During the preparation of this work, the authors used ChatGPT-5 in order to: translate text; paraphrase and reword sentences; improve writing style; and check grammar and spelling errors. After using this tool, the authors reviewed and edited the content as needed and take full responsibility for the publication's content.

\balance{}
\bibliographystyle{ACM-Reference-Format}
\bibliography{references}

\end{document}